\documentclass[twocolumn,aps,showpacs,preprintnumbers,amsmath,amssymb,unsortedaddress,superscriptaddress]{revtex4}

\usepackage{graphicx,epsf,epsfig,ulem}
\usepackage{bm}
\usepackage{subfigure}
\usepackage{color}

\begin{document}

\title{Probing electronic excitations in molecular conduction}

\author{B. Muralidharan}
\affiliation{School of Electrical and Computer Engineering, Purdue University,
West Lafayette, IN 47907}
\author{A. W. Ghosh}
\affiliation{Dept. of Electrical and Computer Engineering, University of Virginia,
Charlottesville, VA 22903}
\author{S. Datta}
\affiliation{School of Electrical and Computer Engineering, Purdue University,
West Lafayette, IN 47907}

\date{\today}
 \begin{abstract}
We identify experimental signatures in the current-voltage (I-V)
characteristics of weakly contacted molecules directly arising
from excitations in their many electron spectrum. The current is
calculated using a multielectron master equation in the Fock
space of an exact diagonalized model many-body Hamiltonian for a
prototypical molecule. Using this approach, we explain several
nontrivial features in frequently observed I-Vs in terms of a rich
spectrum of excitations that may be hard to describe adequately
with standard one-electron self-consistent field (SCF) theories.
\end{abstract}
\maketitle

{Theoretical efforts to model molecular conduction have largely
been based on SCF models for electron-electron
interactions \cite{rdatta}}. While they have been fairly successful in
describing both shapes and magnitudes of various I-V characteristics
\cite{rmrs,rasymm}, notable exceptions include low-temperature
measurements on unconjugated and weakly coupled molecules
\cite{jpark,rweber1,rweber2,pnas}, as well as short conjugated molecules
\cite{rreed} where there are clear disagreements between theory and
experiment. Some disagreements could be attributed to uncertainties in
geometry or parasitic resistances; nevertheless the applicability of SCF
approaches need to be scrutinized, especially in the weak coupling
regime. Charging energies of short molecules ($\sim 3$ eV for benzene)
are often larger than their contact induced broadenings ($\le 0.2$ eV for
benzene dithiol on gold), while couplings between various molecular
units ($\sim 2$ eV for conjugated molecules, much less for
non-conjugated species) can be tuned widely using synthetic chemistry.
It is thus debatable whether a molecule acts as a {\it{quantum wire}}
in the SCF regime, or as a {\it{quantum dot}} in the Coulomb Blockade
(CB) regime.
\begin{figure}[ht]
\centerline{\epsfxsize=3.2in\epsfbox{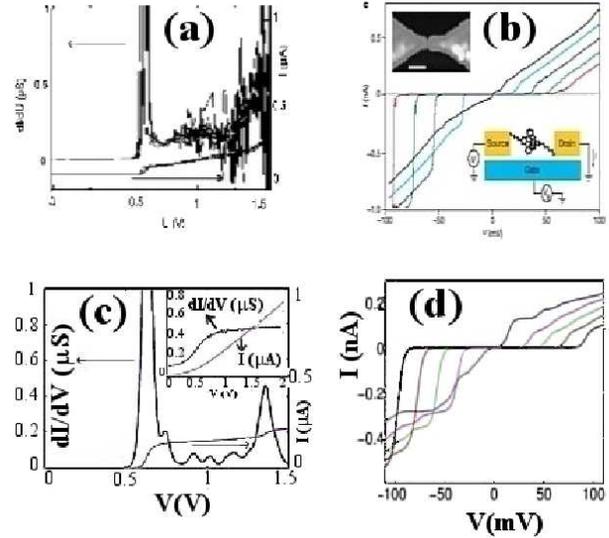}} 
\caption{(color online) (a, b) Experimental \cite{rweber1, jpark} and
(c, d) theoretical I-Vs for a molecular ring weakly coupled with a backbone or with conducting electrodes. 
Many nontrivial, features such as low zero-bias conductance, sharp current onset,
and a subsequent quasilinear region spanning several volts with
multiple closely spaced features (a-d) arise in our treatment of CB
from excitations, but not even qualitatively in a spin restricted SCF
(RSCF) treatment for the same parameter set \cite{cav2} (inset in c). For
asymmetric contacts, there are additional features (b, d) including
current step heights (as opposed to widths) that are asymmetric in
bias, are modulated with a gate voltage \cite{jpark}, and
reverse polarity for gate voltages on either side of the charge
degeneracy point \cite{rscott}.}
\label{f0}
\end{figure}

In this paper, we employ a multielectron master equation
\cite{rralph,rhettler} in the Fock space of a prototypical molecular
Hamiltonian to describe conduction through molecules with weak contact
couplings or poor conjugation. A full many-body treatment of transport
even with a small molecule, modeled simply as an array of quantum dots,
yields many features with compelling similarities (Fig.~1) to relevant
experiments \cite{rweber1,rweber2,jpark}. These features, however, are
quite difficult to obtain using a traditional non-equilibrium Green's
function (NEGF) treatment of transport, being only {\it{perturbative}}
in the interaction parameter \cite{rjauho}. A spin restricted (RSCF)
calculation (inset in Fig. 1c) typically creates slow current onsets
spread over several volts by Coulomb costs for adiabatic charging. The
high zero-bias conductances, in clear variance with experiments, could
be removed by incorporating self-interaction correction for integer
charge addition in the CB regime. However, crucial to experiments in
this regime is the fact that the molecule can also execute transitions
between various {\it{excited states}} of the neutral and singly charged
species {\it{at no additional Coulomb cost}}, making it possible to
directly probe a rich spectrum of such transition levels within a small
bias window.

It seems difficult to capture this rich spectrum adequately within any
SCF theory even with self interaction correction
\cite{rpal,ssan,rdatta2} or effective one electron potentials
\cite{rdelaney}, especially under non-equilibrium conditions.  A single
spin-degenerate level (Fig. 2) illustrates the problem.  While the
deficiencies of SCF (e.g.  adiabatically smeared steps) are
rectified with self-interaction corrections using a spin-unrestricted
calculation for {\it{equilibrium}} properties such as $N-\mu$ (Fig.
2b), the same approach gives the wrong {\it{nonequilibrium}} properties
such as current step heights. An unrestricted calculation yields
equal step heights for each spin removal, while the exact result using
rate equations predicts that the first step is two-thirds of the second
(Fig. 2c), there being two ways to remove (add) the first
spin for a filled (empty) level, but only one way to remove (add) the
second one (Fig 2a).

\begin{figure}
\centerline{\epsfig{figure=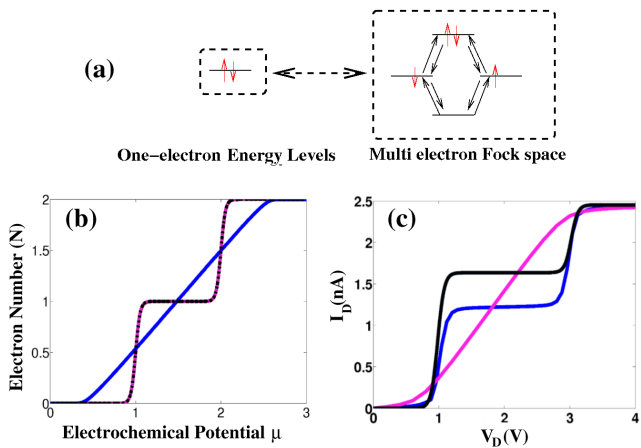,width=3.4in}}
\caption{(color online) (a) Fock space, (b) equilibrium occupancy
$N-\mu$ and (c) nonequilibrium I-V of a spin degenerate level $\epsilon=1eV$ 
with a single electron charging energy $U=1eV$. A restricted SCF (RSCF) 
(pink) calculation shows fractional charge
occupation and is inappropriate in the weak coupling limit. A Spin
Unrestricted SCF (blue) describes integer charge transfer and matches
the many-body $N-\mu$ (black) plot; however, it yields {\it{equal
current steps}} corresponding to sequential removal of two electrons,
as opposed to a many-body calculation (shown in black) in which step
heights are in the ratio of 2:1. Including correlations in SCF alters
the current onsets and the plateau widths, but misses the essential
point that {\it{consecutive removal of spins need not carry equal
current}} \cite{rralph,rralph2}.}
\label{f1}
\end{figure}

The discrepancy with SCF becomes more pronounced for multiple orbitals
where a spin can be removed by one contact from the ground state and
reinjected by the other into feasible excitations of the neutral and
singly charged systems, creating additional features within the Coulomb
Blockade plateaus. Such excitations, crucial to the experiments
addressed here, arise {\it{non-perturbatively}} from our rate equations
through exact diagonalization of the many-body Hamiltonian, going
beyond orthodox Coulomb Blockade theory \cite{rlikharev} due to size
quantization and transitions among discrete many-body states. As
the size of multielectron Fock space increases exponentially with the
number of basis functions, we employ a minimal basis set in a reduced
single-particle Hilbert space that captures the conjugation chemistry
and electrostatics and yet allows exact diagonalization \cite{cav2}. Quantitative
justice to chemistry would possibly require looking at a reduced subset
of excitations (partial configuration interaction \cite{rdelaney2})
within a multiorbital description. Our aim is to solve the transport
problem exactly for a simple system, rather than do an approximate SCF
calculation of a more elaborate quantum chemical system \cite{caveat}.

{\it{Approach.}} We start with a tight-binding model for benzene (one
orbital per atom), with onsite, hopping, and Hubbard parameters that
can be benchmarked with separate LDA calculations \cite{rbhasko}. In
contrast with single dot studies, long-ranged Coulomb terms (modeled
with the Mataga-Nishimoto approximation \cite{rmagnus}) and hopping are
responsible for off-diagonal correlations in the charging term of the
molecular eigenspace. Exact diagonalizing this Hamiltonian yields a
large spectrum of closely spaced excitations in every charged
configuration of the molecule.  Using the equation of motion of the
density matrix of the composite molecule and leads and assuming no
molecule-lead correlations, one can derive \cite{timm,rbraig} a simple
master equation for the density-matrix of the system.  Ignoring
off-diagonal coherences, we are left with a master equation
\cite{rbraig} in terms of the occupation probabilities $P^N_i$ of each
N electron many-body state $|N,i\rangle$ with total energy $E^N_i$. The
master equation then involves transition rates
$R_{(N,i)\rightarrow(N\pm 1,j)}$ between states differing by a single
electron, leading to a set of independent equations defined by the size
of the Fock space \cite{rralph}
\begin{equation}
\frac{dP^N_i}{dt} = -\sum_j\left[R_{(N,i)\rightarrow(N\pm 1,j)}P^N_i -
R_{(N\pm 1,j)\rightarrow(N,i)}P^{N\pm 1}_j\right]
\label{ebeenakker}
\end{equation}
along with the normalization equation $\sum_{i,N}P^N_i = 1$. For weakly
coupled dispersionless contacts, parameterized using bare-electron
tunneling rates $\gamma_{\alpha}$, ($\alpha$: left/right contact), we
define rate constants
\begin{eqnarray}
\Gamma_{ij\alpha}^{Nr} &=& \gamma_\alpha|\langle N,i|c^\dagger_\alpha|N-1,j\rangle|^2\nonumber\\
\Gamma_{ij\alpha}^{Na} &=& \gamma_\alpha|\langle N,i|c^{}_\alpha|N+1,j\rangle|^2,
\end{eqnarray}
$c^\dagger_\alpha,c^{}_\alpha$ are the creation/annihilation operators for an electron on the
molecular end atom coupled with the corresponding electrode.
The transition rates are given by
\begin{eqnarray}
R_{(N,i)\rightarrow(N-1,j)} &=&
\sum_{\alpha=L,R}\Gamma_{ij\alpha}^{Nr}\left[1-f(\epsilon^{Nr}_{ij}-\mu_\alpha)\right]
\nonumber\\
R_{(N-1,j)\rightarrow(N,i)} &=&
\sum_{\alpha=L,R}\Gamma_{ij\alpha}^{Nr}f(\epsilon^{Nr}_{ij}-\mu_\alpha).
\end{eqnarray}
for the removal levels $(N,i \rightarrow N-1,j)$, and replacing $(r
\rightarrow a,   f \rightarrow 1-f)$ for the addition levels $(N,i
\rightarrow N+1,j)$. $\mu_\alpha$ are the contact electrochemical
potentials, $f$ is the corresponding Fermi function, with single
particle removal and addition transport channels $\epsilon^{Nr}_{ij} =
E^N_i - E^{N -1}_j$, and $\epsilon^{Na}_{ij} = E^{N+1}_j - E^N_i$.
Finally, the steady-state solution to Eq.(\ref{ebeenakker}) is used to
get the left terminal current
\begin{equation}
I = \pm\frac{e}{\hbar}\sum_{ij}\left[R^L_{(N,i)\rightarrow(N\pm 1,j)}
P^N_i - R^L_{(N\pm 1, j)\rightarrow(N,i)}P^{N\pm 1}_j \right]
\end{equation}
where states corresponding to a removal of electrons by the left
electrode involve a negative sign.

\begin{figure}
\centerline{\epsfxsize=3.2in\epsfbox{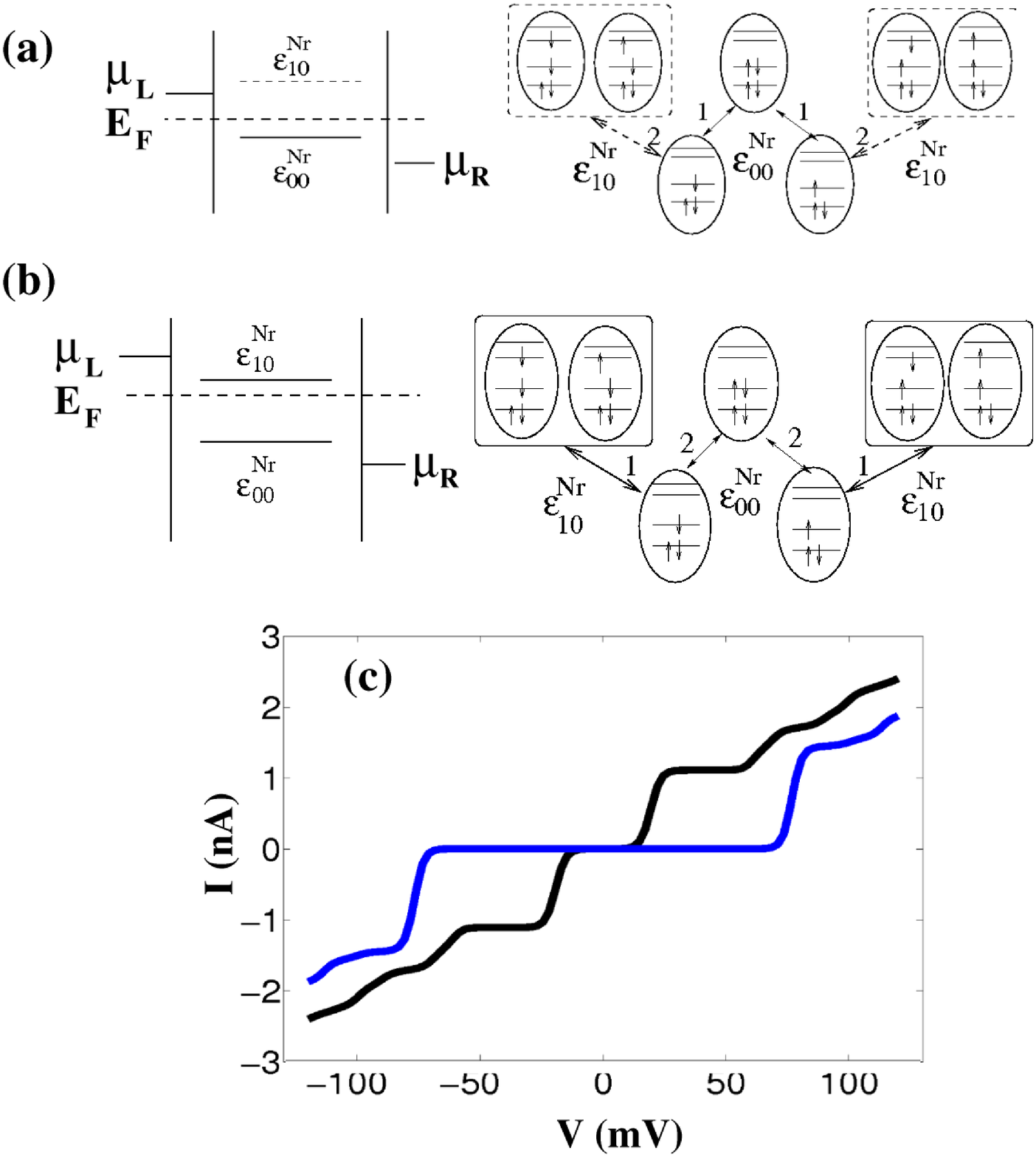}}
\caption{(color online) Coulomb Blockade I-V features for a general
molecular system, Threshold transition involving 1) only ground
states. Here $|E_F-\epsilon^{Nr}_{00}| =10$ meV:  
$\epsilon^{Nr}_{00}$ is accessed before $\epsilon^{Nr}_{10}$ (shown in the adjacent state transition diagram). 
I-V characteristics (shown black in (b)) then has a brief intervening plateau until an
excitation is accessed. 
2) Excited states:  Conduction at threshold ($|E_F-\epsilon^{Nr}_{00}| =30$ meV)
involves a transport channel involving excited states also (say
$\epsilon^{Nr}_{10}$) i.e., $\epsilon^{Nr}_{10}$ is accessed before
$\epsilon^{Nr}_{00}$. In this case (see text) current rise (shown blue
in (c)) due to closely spaced excitations follows upon threshold.}
\label{f2}
\end{figure}

{\it{Results.}} We calculate current in a break-junction configuration
with equal electrostatic coupling with the leads, setting $\mu_{L,R} =
E_F \mp eV_d/2$, and equal resistive couplings set by the
ratio $\gamma = \gamma_L/\gamma_R = 1, \gamma_L=0.6$ meV. Coulomb
Blockade with integer charge transfer manifests itself as a vanishing
pre-threshold current followed by a step wise increase in current
\cite{jpark,rscott,rweber1,rweber2}. The onset of conduction is
established by the offset between the equilibrium Fermi energy $E_F$
and the first accessible transition energy (focussing on removal levels
for concreteness, this corresponds to the transport channel marked
$\epsilon^{Nr}_{00}$ in Figs. 3a and 3c). The onset can be varied by
varying the gate voltage, thereby accounting for the variation in
conductance gap \cite{rralph2} with gate bias.

The simplest impact of Coulomb Blockade on the I-Vs of short molecular
wires is a clear suppression of zero-bias conductance, often seen
experimentally \cite{rreed,rtao}.  Indeed, a spin unrestricted SCF with
self-interaction corrections \cite{rpal,ssan} can yield a Coulomb
staircase with intervening plateaus through the Coulomb cost of adding
or removing an electron to the molecular ground state. However, integer
charge transfer can also occur between various electronic
{\it{excitations}} of the neutral and singly charged species at
marginal correlation costs \cite{rfulde}. The above fact leads to fine
structure in the plateau regions \cite{jpark,rweber1,rweber2,pnas},
specifically, a quasilinear regime resulting from very closely spaced
transport channels ($\epsilon^{N}_{ij}$) via excitations. The crucial
step is the access of the first excited state via channel
$\epsilon^{Nr}_{10}$, following which transport channels involving
higher excitations are accessible in a very small bias window. The
sequence of access of transport channels upon bias, enumerated in the
state transition diagrams shown in Figs. 3a and 3b, determines the
shape of the I-V. When the Fermi energy $E_{F}$ lies closer to the
threshold transport channel $\epsilon^{Nr}_{00}$ (Fig. 3a), it takes an
additional positive drain bias for the source to access the first
excited state of the neutral system via the transition
$\epsilon^{Nr}_{10}$, as shown in the state transition diagram in Fig.
3a. The I-V shows a sharp rise followed by a plateau (Fig. 3c), as seen
in various experiments \cite{rdekker}. However, when transport channels
that involve low lying excitations such as $\epsilon^{Nr}_{10}$ are
closer to the Fermi energy $E_F$ than $\epsilon^{Nr}_{00}$ (Fig. 3b),
the excitations get populated by the left contact immediately when the
right contact intersects the threshold channel $\epsilon^{Nr}_{00}$,
allowing for a {\it{simultaneous}} population of both the ground and
first excited states via $\epsilon^{Nr}_{00}$ and $\epsilon^{Nr}_{10}$
at threshold.  Under these conditions the I-V shows a sharp onset
followed immediately by a quasilinear regime (Fig. 3c) with no
intervening plateaus, as observed frequently in I-Vs of molecules
weakly coupled with a backbone \cite{rweber1,rweber2,jpark}.

\begin{figure}
\centerline{\epsfig{figure=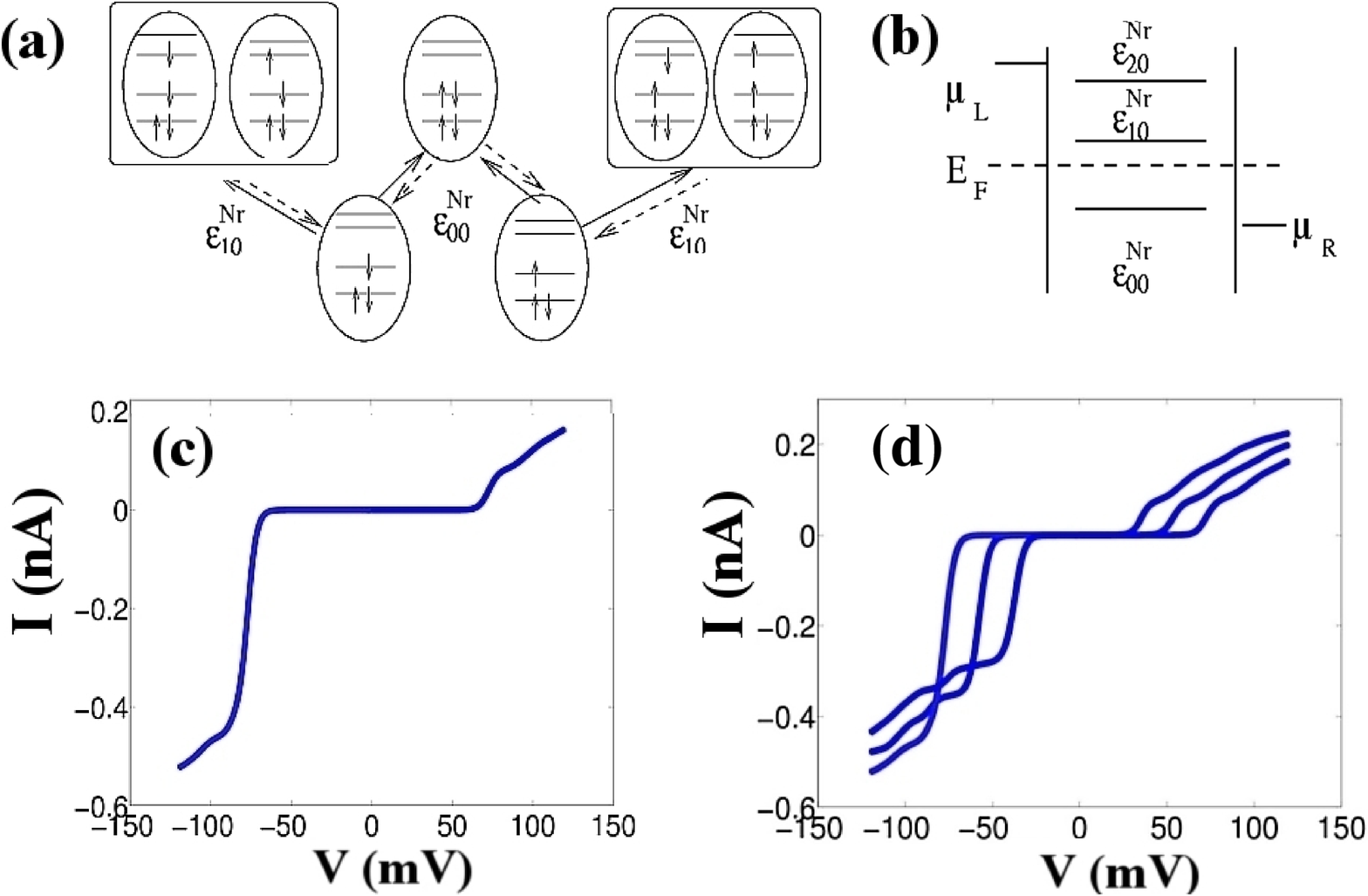,width=3.4in,height=2.4in}}
\caption{(a) State transition diagram showing various addition and
removal pathways for asymmetric contacts ($\gamma_L \gg \gamma_R$),
including the possibility of populating higher excitations (b), say,
via transport channel $\epsilon^{Nr}_{20}$ at threshold.  For positive
bias charge removal is the rate limiting process, while for negative bias
addition dominates, accounting thus for the corresponding I-V asymmetry
in (c). Progressive access of higher excitations also accounts for the
observed gate modulation of the current steps, as shown in (d).}
\label{f3}
\end{figure}

The direct role of excitations in conduction becomes particularly striking
under asymmetric coupling ($\gamma=100,\gamma_L=0.6$ meV) with contacts \cite{jpark,rscott}.
In contrast to the SCF regime where unequal charging drags out the same
level current over different voltage widths \cite{rasymm}, in the CB
regime the current step heights  themselves are asymmetric at threshold
(Fig. 4c). This asymmetry arises due to the difference in the number
of pathways for removing or adding a spin, also taking into account the
possible excitation channels between the neutral and singly charged
species (Fig. 4 a, b).  The number of such accessible excitations
at threshold can be altered with an external gate bias, leading to
a prominent gate modulation of the threshold current heights, over
and above the modulation of the onset voltages and the conductance
gap \cite{jpark} (Fig. 4d). Furthermore, it is easy to show that the
asymmetry will flip between gate voltages on either side of the charge
degeneracy point, as is also observed experimentally \cite{rscott}.
While the qualitative features of our I-Vs are robust with respect to
variation of our model parameters, details specific to experiments
(e.g. onset voltages, polarization asymmetries \cite{rweber1} and
temperature dependences \cite{rweber2}) can vary and will be discussed in
detail elsewhere \cite{rbhasko}. For instance, correlation alone cannot
explain ultralow peak currents through a level since those depend only
on contact couplings through the ratio $\gamma_L\gamma_R/(\gamma_L +
\gamma_R)$. This predicts a peak current $\sim$ 3 $\mu A$ for a $0.1 eV$
broadening as in chemisorbed benzene dithiol \cite{rdelaney2}, still
much larger than experiments \cite{rreed}, indicating that one needs
further to postulate weak couplings due to non-ideal contact couplings
or perhaps parasitic resistances from multiple molecules \cite{rkircz}.
Further complications could arise from strong electron-phonon interactions
\cite{jpark} that smoothen out the first few conduction plateaus (Fig.~\ref{f0}d) 
due to low lying excitations 
over phonon energies significantly smaller than their
Coulomb counterparts at tens of meV.

In summary, we have used a rate equation in the Fock space of a molecular
Hamiltonian to address significant experimental features like
suppressed zero-bias conductances, sharp steps that are often asymmetric,
gate modulated and interchangeable, and followed occasionally by
extended quasiohmic regimes. While our method is particularly suited to
systems with large charging and small coupling, the opposite regime is
usually handled perturbatively by SCF-NEGF. Developing the transport
formalism for the intermediate coupling regime could be nontrivial
\cite{rgurevich}, involving novel physics due to the interplay between
charging (localization) and hybridization (delocalization), and may
be crucial to understanding a variety of other molecular switching and
sensing-based phenomena already being explored experimentally.

We would like to thank S. Pati, M. Reed, G. Klimeck, M. Korkusinski, D.
Kienle and E.  Polizzi for useful discussions. This work was funded by
DURINT and DARPA-ONR.


\begin{thebibliography}{100}
\bibitem{rdatta} S. Datta {\it{et al.}}, Phys. Rev. Lett. {\bf{79}},
2530 (1997); P. S. Damle {\it{et al.}}, Chem. Phys. {\bf{281}}, 171
(2002); M. Di Ventra {\it{et al.}}, Phys. Rev. Lett. {\bf{84}}, 979
(2000); J. Taylor {\it{et al.}}, Phys. Rev. B {\bf{63}}, 245407
(2001).

\bibitem{rmrs} A. W. Ghosh {\it{et al.}}, MRS Bull. {\bf{29}},
391 (2004).

\bibitem{rasymm} F. Zahid {\it{et al.}}, Phys. Rev. B {\bf{24}},
245317 (2004).


\bibitem{jpark} J. Park {\it{et al.}}, Nature {\bf{417}}, 722 (2002).

\bibitem{rweber1} J. Reichert {\it{et al.}}, Appl. Phys. Lett. {\bf{82}},
4137 (2003).

\bibitem{rweber2} M. Mayor, Angewandte Chemie Int. Ed. {\bf{42}}, 5834
(2003).

\bibitem{pnas} M. Elbing {\it{et al.}}, Proc. Natl. Acad. Sc. {\bf{102}},
8915 (2005).

\bibitem{rreed} M. A. Reed {\it{et al.}}, Science {\bf{278}}, 252 (1997).

\bibitem{rralph} E. Bonet {\it{et al.}}, Phys. Rev. B {\bf{65}}, 045317
(2002); C. W. J. Beenakker, Phys. Rev. B {\bf{44}}, 1646 (1991).

\bibitem{rhettler} M. H. Hettler {\it{et al.}}, Phys. Rev. Lett.
{\bf{90}}, 076805 (2003).

\bibitem{rjauho} `Quantum Kinetics in Transport and Optics of Semiconductors' H. Haug, and A.-P. Jauho,
 Springer Series in Solid-State Sciences 123, Springer-Verlag Berlin Heidelberg, 1996.

\bibitem{rscott} G. Scott {\it{et al.}}, cond-mat/0405345.

\bibitem{rpal} J. J. Palacios, cond-mat/0505565.

\bibitem{ssan} C. Toher {\it{et al.}}, cond-mat/0506244.

\bibitem{rdatta2} S. Datta, Nanotechnology {\bf{15}}, S433 (2004).

\bibitem{rdelaney}
A. Ferretti {\it{et al.}}, cond-mat/0409222; Na Sai {\it{et
al.}}, cond-mat/0411098; K. Burke {\it{et al.}}, cond-mat/0502385.

\bibitem{cav2}The features explained in the experiments are qualitatively
independent of the quantum chemical model. For Fig 1c we have used in
our prototype molecular Hamiltonian, single electron charging $U=2 eV$
and a level separation $\Delta\epsilon=0.9 eV$, while for calculations
on porphyrin based molecule (Fig. 1d) a charging energy of $U=100$ meV
and $\Delta\epsilon=50$ meV were used.

\bibitem{rralph2} M. M. Deshmukh {\it{et al.}}, Phys. Rev. B {\bf{65}}, 073301 (2002).

\bibitem{rlikharev} `Single Charge Tunneling', Ed. H. Grabert, and M. H. Devoret, NATO ASI series 294,
Plenum Press, New York, 1992.

\bibitem{rdelaney2} P. Delaney and J. C. Greer, Phys. Rev. Lett {\bf{93}}, 036805 (2004).

\bibitem{caveat} While DFT when applied to total energy
calculations is not a mean field or SCF theory, its applications to transport problems
so far has been in a mean field sense.
\bibitem{rbhasko} B. Muralidharan {\it{et al.}}, in preparation.

\bibitem{rmagnus} M. Paulsson and S. Stafstr\"{o}m, Phys. Rev. B {\bf{64}},
035416 (2001).

\bibitem{timm} F. Elste and C. Timm, Phys. Rev. B {\bf{71}}, 155403 (2005).

\bibitem{rbraig} S. Braig and P. W. Brouwer, cond-mat/0412592.

\bibitem{rtao} X. Xiao {\it{et al.}}, Nano Lett. {\bf{4}}, 267 (2001);
B. Q. Xui {\it{et al.}}, Science {\bf{301}}, 1221 (2003).

\bibitem{rfulde} `Electron Correlations in Molecules and Solids', P. Fulde,
Springer Series in Solid-State Sciences 100, Springer-Verlag Berlin Heidelberg,
1991.

\bibitem{rdekker} J-O. Lee {\it{et al.}}, Nano Lett. {\bf{3}}, 113 (2003).


\bibitem{rkircz} E. Emberly {\it{et al.}}, Phys. Rev. B {\bf{64}},
235412 (2001).

\bibitem{rgurevich} S. A. Gurvitz and Y. S. Prager, Phys. Rev. B
{\bf{53}}, 15932 (1996);  J. N. Pedersen and A. Wacker, cond-mat/0509024.

\end{thebibliography}
\end{document}